\documentclass[3p,times,twocolumn]{elsarticle}

\usepackage{ecrc}

\volume{00}

\firstpage{1}

\journalname{Nuclear and Particle Physics Proceedings}

\runauth{}

\jid{nppp}

\jnltitlelogo{Nuclear and Particle Physics Proceedings}

\usepackage{amssymb}

\usepackage[figuresright]{rotating}

\begin{document}

\begin{frontmatter}



\dochead{}

\title{Resolving the $R_{AA}$ to $v_n$ puzzle}


\author{Jacquelyn Noronha-Hostler}

\address{\small{\it  Department of Physics, University of Houston, Houston, TX 77204, USA}}

\begin{abstract}
After 10 years of struggling to simultaneously describe the nuclear modification factor $R_{AA}$ and flow harmonics $v_n$'s at high $p_T$, now theoretical models are able to reproduce experimental data well. The necessary theoretical developments such as event-by-event fluctuations, choice of initial conditions, and the scalar product method to calculate flow harmonics at high $p_T$ are reviewed.  Additionally, a discussion of new proposed experimental observables known as Soft Hard Event Engineering (SHEE) that are sensitive to the path length dependence of the energy loss is included.  

\end{abstract}


\end{frontmatter}

\section{Introduction}
\label{Introduction}

Relativistic heavy-ion collisions have successfully recreated the Quark Gluon Plasma (QGP) in the laboratory at RHIC and the LHC.  While it is impossible to make real-time observations of its dynamics due to confinement of quarks and gluons, one can work ``back in time" using its signatures to confirm its existence.  Two of the most convincing signatures of the QGP are (nearly) perfect fluidity and jet suppression.  

Perfect fluidity arises around the strongly interacting cross-over phase transition \cite{Aoki:2006we} from the QGP into the Hadron Gas Phase \cite{Danielewicz:1984ww,Kovtun:2003wp,NoronhaHostler:2008ju,NoronhaHostler:2012ug}. Strong evidence for perfectly fluidity comes from the enormous success of event-by-event relativistic viscous hydrodynamical models in describing collective flow observables with an extremely small shear viscosity to entropy density ratio \cite{Gardim:2012yp,Gale:2012rq,Niemi:2015qia,Noronha-Hostler:2015uye,Bernhard:2016tnd}.  Elliptical flow, $v_2$, indicates that there was a dominating almond shape in the impact region where two heavy-ion collided while a non-zero triangular flow, $v_3$, arises due to quantum fluctuations of the positions of the nucleons, which can produce a wide variety of initial shape variations around the dominating almond shape \cite{Alver:2010gr}. 

\begin{figure*}[ht]
\centering
\includegraphics[width=0.8\textwidth]{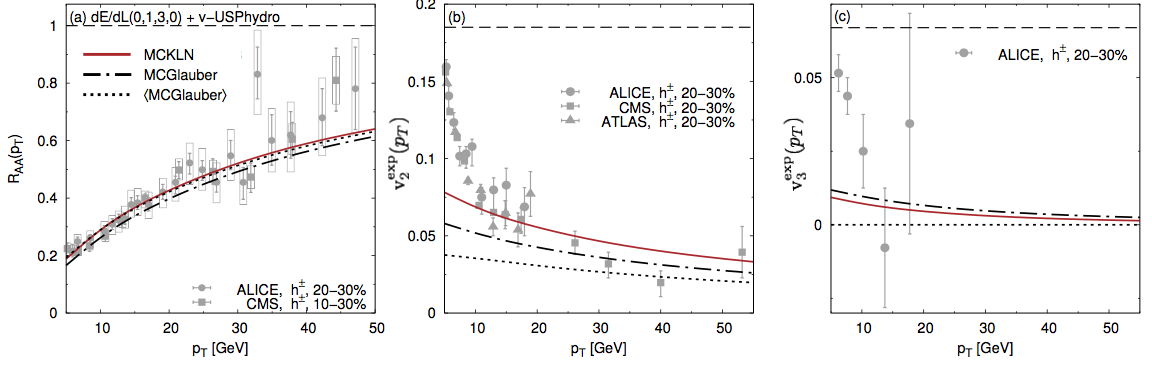}
\caption{(Color online) Model calculations for (a) 
$\pi^0$ $R_{AA}(p_T)$, (b) $v_2\{2\}(p_T)$, (c) $v_3\{2\}(p_T)$ for $20-30\%$ centrality at $\sqrt{s}=2.76$ TeV Pb+Pb collisions at the LHC \cite{Abelev:2012hxa,CMS:2012aa,Abelev:2012di,Chatrchyan:2012xq,Aad2012330}. MCKLN initial conditions are shown in solid red, dotted-dashed black line is for MCGlauber, the black dotted line $\langle{\rm MCGlauber}\rangle$ neglects initial state fluctuations. }
\label{fig:RAAv2v3}
\end{figure*}
Jet suppression uses the fact that hard scattering processes that occur immediately after the collision produce highly energetic jets. In the presence of a strongly interacting dense medium such as the QGP these high momentum particles can lose energy and momentum  \cite{Bjorken:1982tu} and the amount of energy loss is strongly correlated with the path length that the high momentum jet travels across the plasma. Thus, one can imagine that jets produced in an eccentric event would either be highly suppressed along the long axis or still maintain most of its energy along the short axis.  From this understanding, it is natural to normalize the number of high $p_T$ particles in heavy ion collisions to those produced in $pp$ collision times the number of collisions $N_{\rm coll}$, which is known as the nuclear modification factor $R_{AA}=\frac{dN_{AA}/dydp_T d\phi}{N_{\rm coll}\,dN_{pp}/dydp_T}$ \cite{Gyulassy:1990ye,Wang:1991hta,Wang:1991xy}.  Thus, a suppression is seen at high $p_T$ i.e. $R_{AA}<1$, which has historically been well-reproduced by various theoretical energy models \cite{Burke:2013yra}.

Around 10 years ago, a seminar paper with the measurement of high $p_T$ $v_2$ was published \cite{Adler:2005rg}, which was a major step forward towards merging these two signatures of the QGP.  However, a simultaneous description of $R_{AA}$ and $v_2$ was notoriously difficult. In fact, $R_{AA}$ could be reasonably described but the computed $v_2$ underpredicted the data (see, for instance, the discussions in \cite{Betz:2014cza,Xu:2014tda}). While there has always been an understanding that the QGP background affects high momentum particles, it was not until earlier this year that the influence of event-by-event fluctuations and the corresponding initial eccentricities on high momenta particles were studied in detail \cite{Noronha-Hostler:2016eow}. Once event-by-event fluctuations and more realistic initial conditions were implemented, the decade old $R_{AA}\otimes v_2$ was solved \cite{Noronha-Hostler:2016eow}. Additionally, it was found in \cite{Noronha-Hostler:2016eow,Betz:2016ayq} that the high $p_T$ flow harmonics are strongly connected to the initial state eccentricities so one necessary constraint is that the hydrodynamical backgrounds used for energy loss should also reproduce the soft physics flow harmonics as well.  Furthermore, in the heavy flavor sector a connection between the initial state and the heavy flavor $v_2$ is also seen \cite{Nahrgang:2014vza,Nahrgang:2016wig,Prado:2016xbq,Prado:2016szr}.

In this proceedings, the most important advances needed to solve the $R_{AA}\otimes v_2$ puzzle are reviewed.  Additionally, one of the most significant findings in the aftermath of the $R_{AA}\otimes v_2$ puzzle is that event shape engineering can be explored in the high $p_T$ region in order to distinguish between different energy loss mechanisms.

\section{Calculating Flow Harmonics at High $p_T$}
\label{Calculating Flow Harmonics at High $p_T$}

On a more technical note, most experimental measurements of flow harmonics no longer use the event-plane method due to its ambiguous comparisons between theory and experiments but rather the scalar product is used (see \cite{Luzum:2012da}).   In order to calculate the scalar product $v_2\{SP\}$ (or let us call it $v_2\{m\}(p_T)$ where $m$ indicates the number of particles correlated to calculate the flow harmonic) only one high $p_T$ can be used due to the low statistics of high $p_T$ particles and that one high $p_T$ is then correlated with 1 soft particle for $v_2\{2\}(p_T)$ or 3 soft particles for $v_2\{4\}(p_T)$ (see \cite{Zhou:2014bba} for a further discussion). The theoretical analog of $v_n\{2\}(p_T)$ \cite{Noronha-Hostler:2016eow} is then
\begin{equation}\label{eqn:vncor}
v_n\{2\}(p_T)=\frac{\langle v_n\,v_n^{hard}(p_T)\cos\left[n\left(\psi_n-\psi^{hard}_n(p_T)\right]\right) \rangle}{\sqrt{\left\langle \left(v_n\right)^{2}\right\rangle}},
\end{equation}
where $v_n$ is the $n^{th}$ Fourier harmonic of the soft spectra and $v^{hard}_n$ is the $n^{th}$ Fourier harmonic of the particle distribution at high $p_T$. Thus, by its very nature, any high $p_T$ flow harmonic must be a soft-hard correlation and one can intuitively understand the strong connection between soft and hard physics. 

In the experiment, $Q_n$ vectors are used to calculate $v_n\{2\}(p_T)$ on an event-by-event basis \cite{Bilandzic:2010jr,Bilandzic:2013kga} but, theoretically, it is possible to compute the flow anisotropy of high $p_T$ particles from $R_{AA}(p_T,\phi)$ due to oversampling of high $p_T$ particles on a single event, which gives equivalent results in comparisons to experimental data \cite{Khachatryan:2016odn}. Theoretical calculations that model jet-medium interactions with only one high $p_T$ particle embedded within an event would then need to also use the $Q_n$ vectors with a rapidity gap to calculate $v_n\{2\}(p_T)$. Finally, experiments also use multiplicity weighing and centrality rebinning to calculate multiparticle cumulants \cite{Bilandzic:2010jr,Bilandzic:2013kga}, which do have up to a $5\%$ effect on high $p_T$ multiparticle cumulants \cite{Betz:2016ayq} as well as some low $p_T$ cumulants \cite{Gardim:2016nrr}. In depth technical details on the calculation of high $p_T$ flow harmonics can be found in \cite{Betz:2016ayq}.
 
\section{Comparisons to Experimental Data and Predictions}
\label{Comparisons to Experimental Data and Predictions}

\begin{figure}[ht]
\centering
\includegraphics[width=0.4\textwidth]{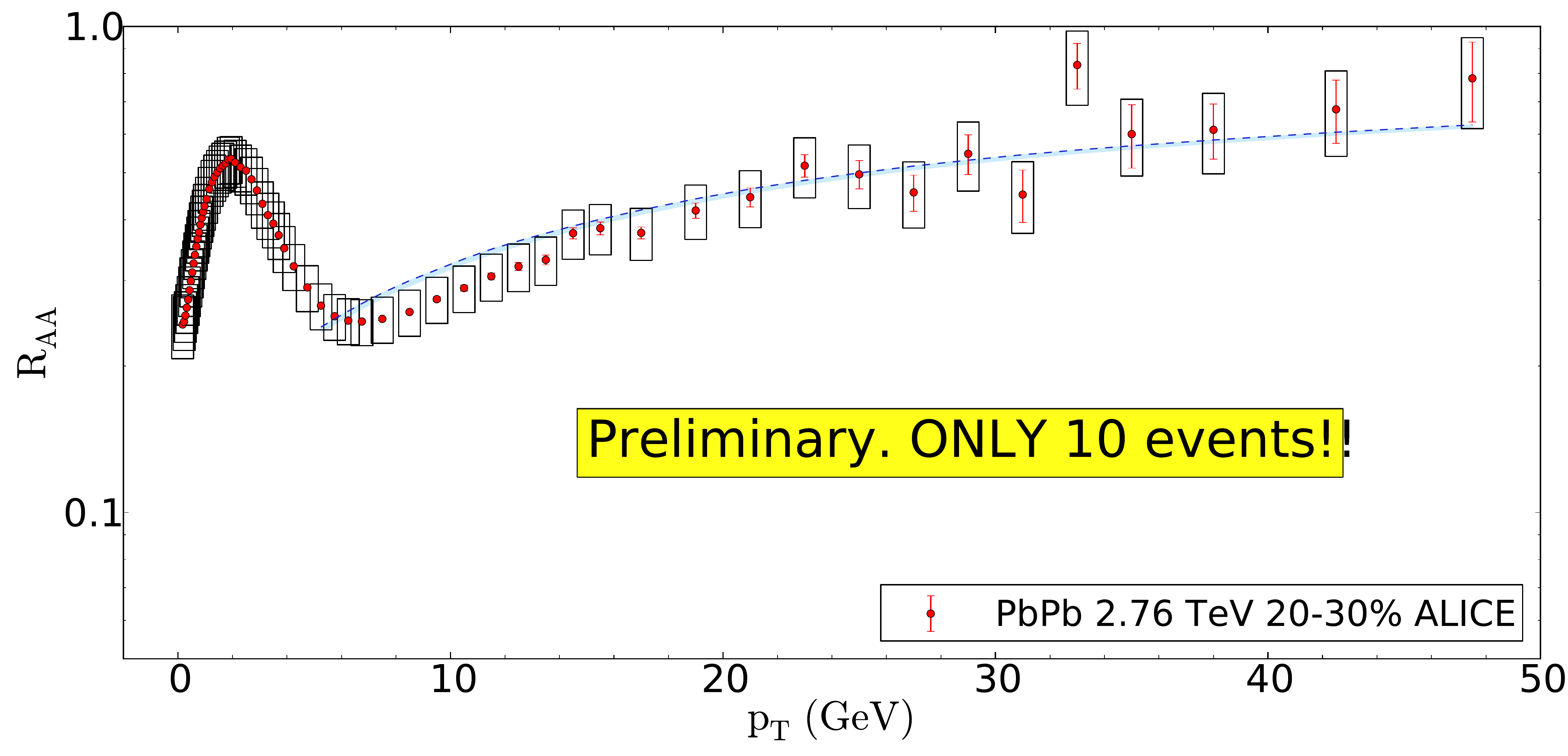}\\
\includegraphics[width=0.4\textwidth]{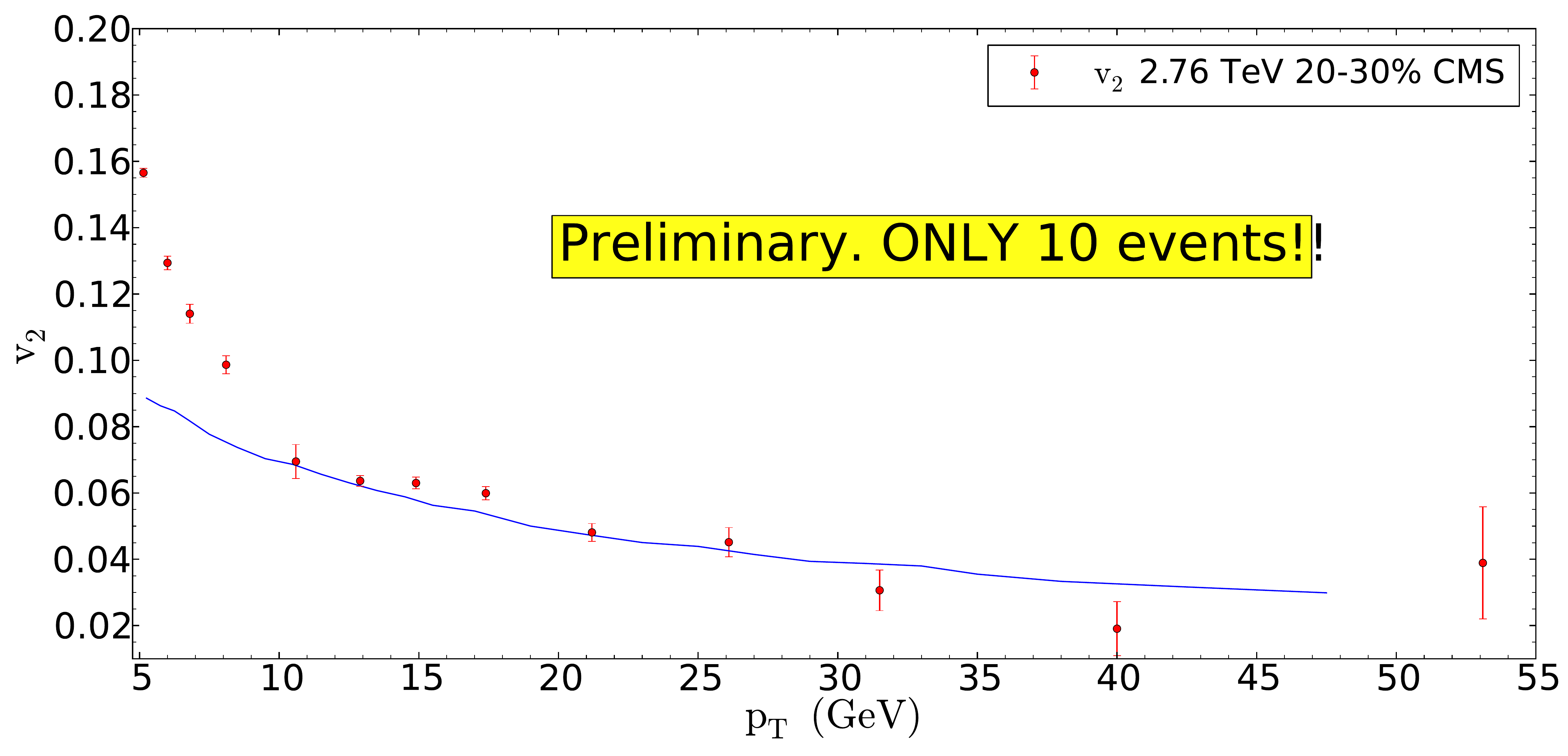}
\caption{(Color online) $R_{AA}$ (top) and $v_2\{SP\}$ (bottom) at LHC PbPb $\sqrt{s_{NN}}=2.76$ TeV  using EKRT initial conditions \cite{Niemi:2015qia} and the quenching weights energy loss mechanism \cite{CasasCarlota,Andres:2016iys}. }
\label{fig:EKRT}
\end{figure}

The first event-by-event $R_{AA}$ to $v_2$ calculations are shown in Fig. \ref{fig:RAAv2v3} using  v-USPhydro+BBMG \cite{Noronha-Hostler:2013gga,Noronha-Hostler:2014dqa,Noronha-Hostler:2015coa,Betz:2011tu,Betz:2012qq}. In Fig. \ref{fig:RAAv2v3} a comparison between two different initial conditions are shown: MCGlauber and MCKLN.  Note that MCKLN has $\sim 30\%$ larger eccentricities, $\varepsilon_2$'s, than MCGlauber \cite{Hirano:2010je,Qiu:2011hf} and in the soft sector it is well-understood that there is a very strong mapping between the initial eccentricities and the final flow harmonics \cite{Gardim:2011xv,Gardim:2014tya}. Thus, it is not surprising that there is also roughly a $30\%$ increase in $v_2\{2\}(p_T)$ as one goes from MCGlauber initial conditions to MCKLN even for high $p_T$.  Indeed, it was shown in \cite{Betz:2016ayq} that there is a very strong linear mapping between $\varepsilon_2$ and $v_2\{2\}(p_T)$ at high $p_T$.

\begin{figure}[ht]
\centering
\includegraphics[width=0.4\textwidth]{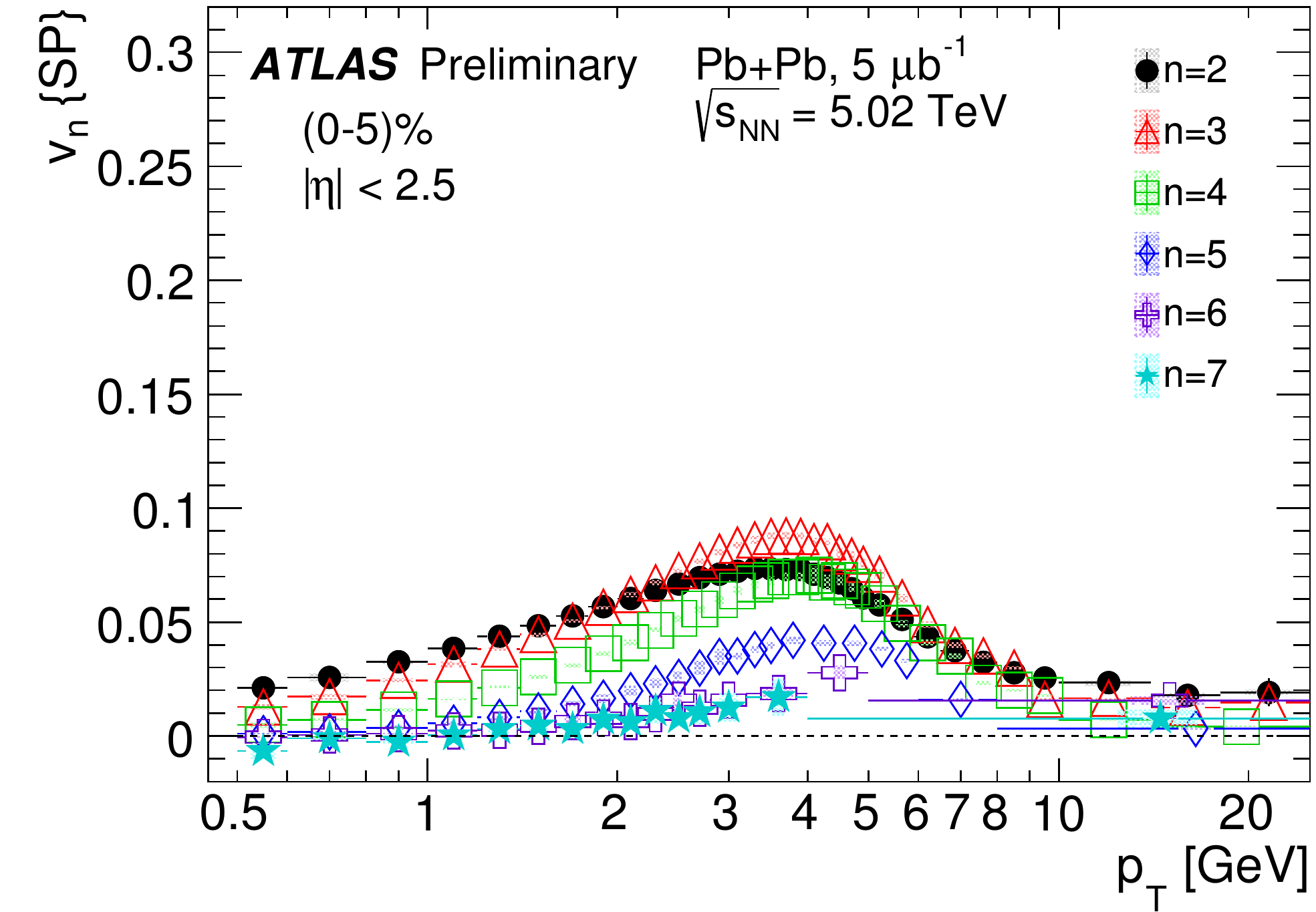} \\
\includegraphics[width=0.4\textwidth]{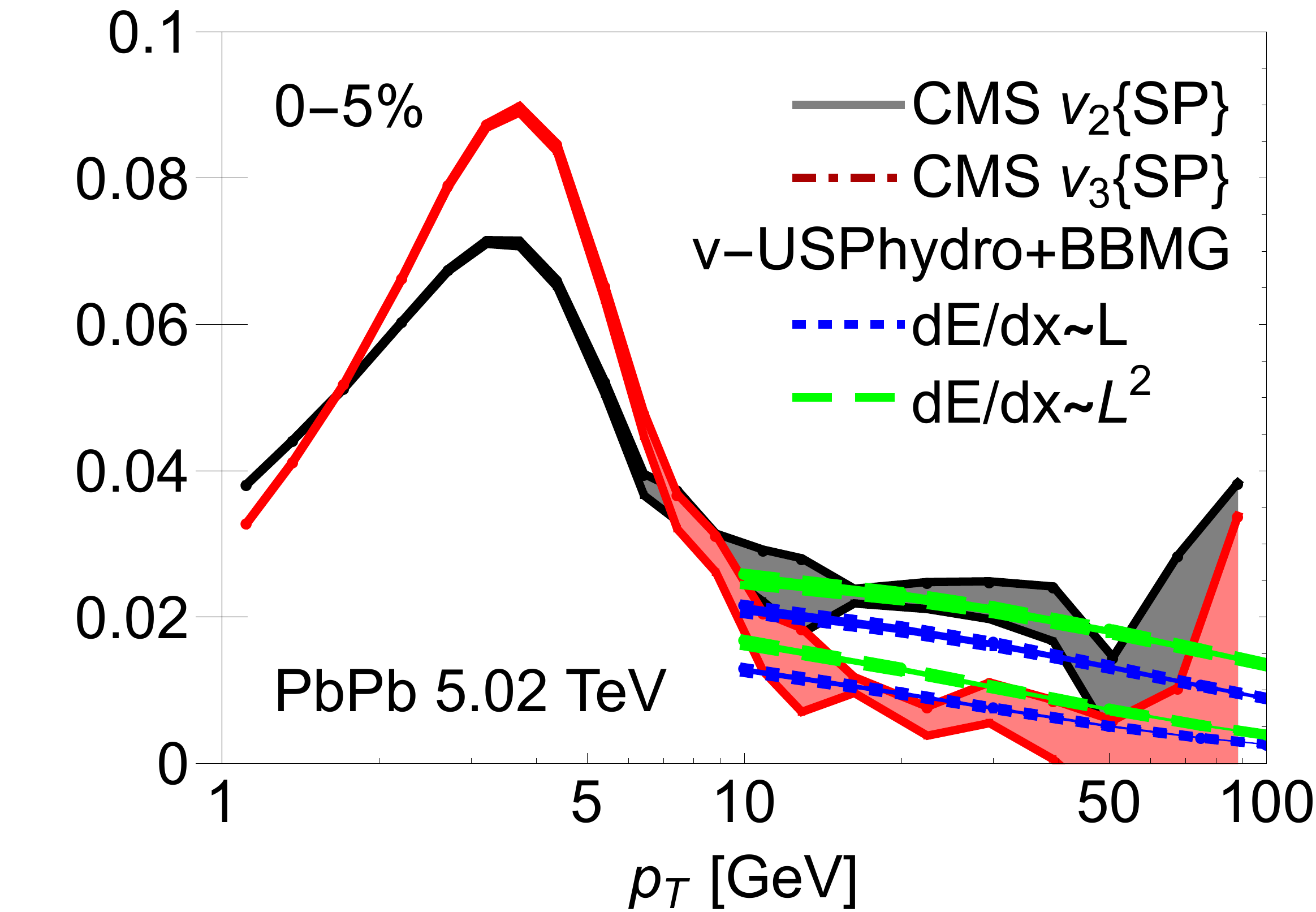} 
\caption{(Color online) $v_2-v_7$ for $0-5\%$ measured up to  $p_T=25$ GeV from ATLAS \cite{ATLAS-CONF-2016-105} (top) and $v_2-v_3$ for $0-5\%$ measured up to  $p_T\sim 100$ GeV from CMS \cite{CMS:2016uwf} compared to predictions from v-USPhydro+BBMG \cite{Betz:2016ayq} for two different energy loss models. }
\label{fig:ATLAS}
\end{figure}

Three clear implications immediately arise from the results in Fig.\ \ref{fig:RAAv2v3}.  The first is that initial conditions should be chosen such that they are able to fit low $p_T$ flow harmonics (see \cite{Noronha-Hostler:2016kjw} for a comparison of MCKLN vs. MCGlauber at low $p_T$).  Indeed, preliminary results using EKRT initial conditions (see Fig. \ref{fig:EKRT}) that fit well soft physics observables \cite{Niemi:2015qia} have already manage to reproduce $R_{AA}$, $v_2\{2\}(p_T)$ and $v_3\{2\}(p_T)$ results at high $p_T$ \cite{CasasCarlota} where the effects on $\hat{q}$ are currently being investigated \cite{Andres:2016iys}. One obvious next step to explore is to reproduce higher order flow harmonics, as measured by ATLAS \cite{ATLAS-CONF-2016-105} using the scalar product method, shown in Fig.\ \ref{fig:ATLAS} (top). 

The second implication is that when one neglects event-by-event fluctuations, one cannot include centrality rebinning/multiplicity weighing, which is always taken into account in the experiment.  Thus, one builds in a systematic bias into the $v_2$ calculation.  However, if one wants to be able to use flow harmonics to distinguish between energy loss models than one could miss the correct physics entirely due to the systematic bias.  In the bottom of Fig\ \ref{fig:ATLAS} comparisons to CMS data are shown for two different energy loss models where only a very small difference is seen between the two and both are roughly within the experimental error bars.  Only using $R_{AA}$ and $v_n$'s across multiple centralities combined with proper treatment of experimental effects can one see a clear difference between energy loss models.    

The third implication is that one can now exploit Soft Hard Event Engineering (SHEE) in order to study energy loss. Significant strides have been made in soft physics studying how different order flow harmonics vary on an event-by-event basis \cite{ALICE:2016kpq} and how soft vs. hard $v_n$'s scale within a centrality class \cite{Aad:2015lwa}.  Suggestions for ways to exploit SHEE are discussed in \cite{Noronha-Hostler:2016eow,Betz:2016ayq,Prado:2016szr,Christiansen:2016uaq}.  

\subsection{Soft Hard Event Engineering (SHEE)}
\label{Soft Hard Event Engineering}

\begin{figure}[ht]
\centering
\includegraphics[width=0.45\textwidth]{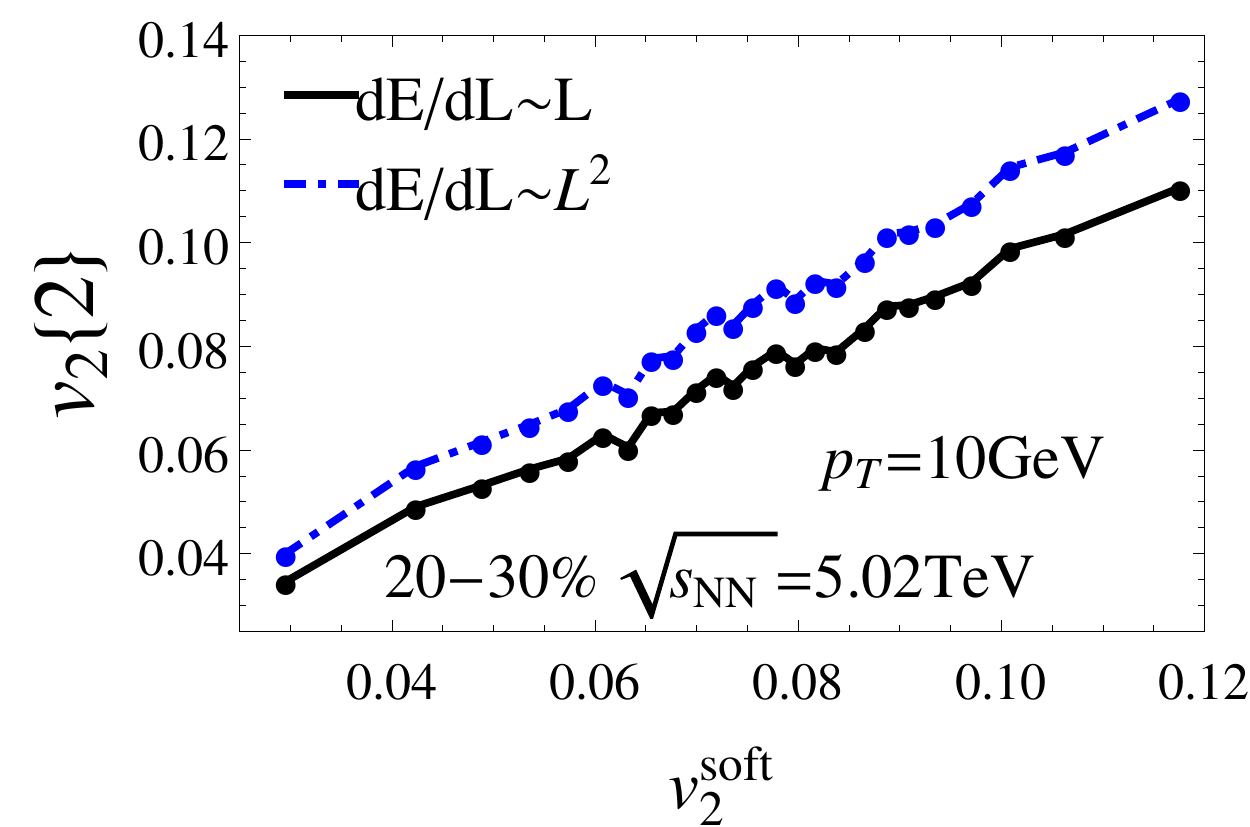} 
\caption{(Color online) Soft-Hard Event Engineering (SHEE) of $v_2$ for $20-30\%$ at $p_T=10$ GeV.}
\label{fig:SHEE}
\end{figure}

Returning to the idea of SHEE of elliptical flow harmonics in \cite{Aad:2015lwa}, within a set centrality class the events are sorted and binned by their integrated (soft) $v_2\{2\}$.  Then, within those bins the respective high $p_T$ $v_2\{2\}$ is also calculated.  If there 
were no high $p_T$ fluctuations of flow harmonics the relationship would be entirely flat.  However, ATLAS data in \cite{Aad:2015lwa} already showed that at $\sqrt{s_{NN}}=2.76$ TeV there is a linear scaling between $v_2^{soft}\{2\}$ and $v_2^{hard}\{2\}$ up to at least $p_T=15$ GeV and it should be possible to to calculate the same quantity up to large $p_T$ at LHC run 2. In Fig.\ \ref{fig:SHEE} SHEE of $v_2$ is shown at $p_T=10$ GeV and it demonstrates a clear splitting depending on the path length dependence of the energy loss.  Such a calculation could be vital in distinguishing between relatively equivalent energy loss mechanisms as shown in \cite{Prado:2016szr} for the heavy flavor sector.

SHEE of flow harmonics provides a much more rigorous test of energy loss mechanisms.  In recent years, the prominence of Monte Carlo generators used to explore jet substructure has received a significant amount of attention \cite{Armesto:2009fj,Young:2011ug,Pang:2012he,Zapp:2013vla,Casalderrey-Solana:2014bpa,Kordell:2016njg}. Transport codes have also successfully reproduced a number of experimental observables in the heavy-flavor sector \cite{Das:2015ana,Esha:2016svw,Senzel:2016qau}. However, event-by-event fluctuations of the initial conditions in this case have largely been ignored.  Including SHEE calculations may be a crucial distinguishing factor between energy loss mechanisms.  For instance, in Fig.\ \ref{fig:SHEE} there is a clear relationship between the path length dependence of the energy loss and the slope of SHEE of $v_2$ where a large power of $n$ in $dE/dx\sim L^n$ produces a steeper slope.  A $dE/dx\sim L^3$ would likely have an even steeper slope in Fig.\ \ref{fig:SHEE}. While not shown here, such a calculation is possible even up to high $p_T$ and is only limited by the error bars of experiments. 

\begin{figure}[ht]
\includegraphics[width=0.5\textwidth]{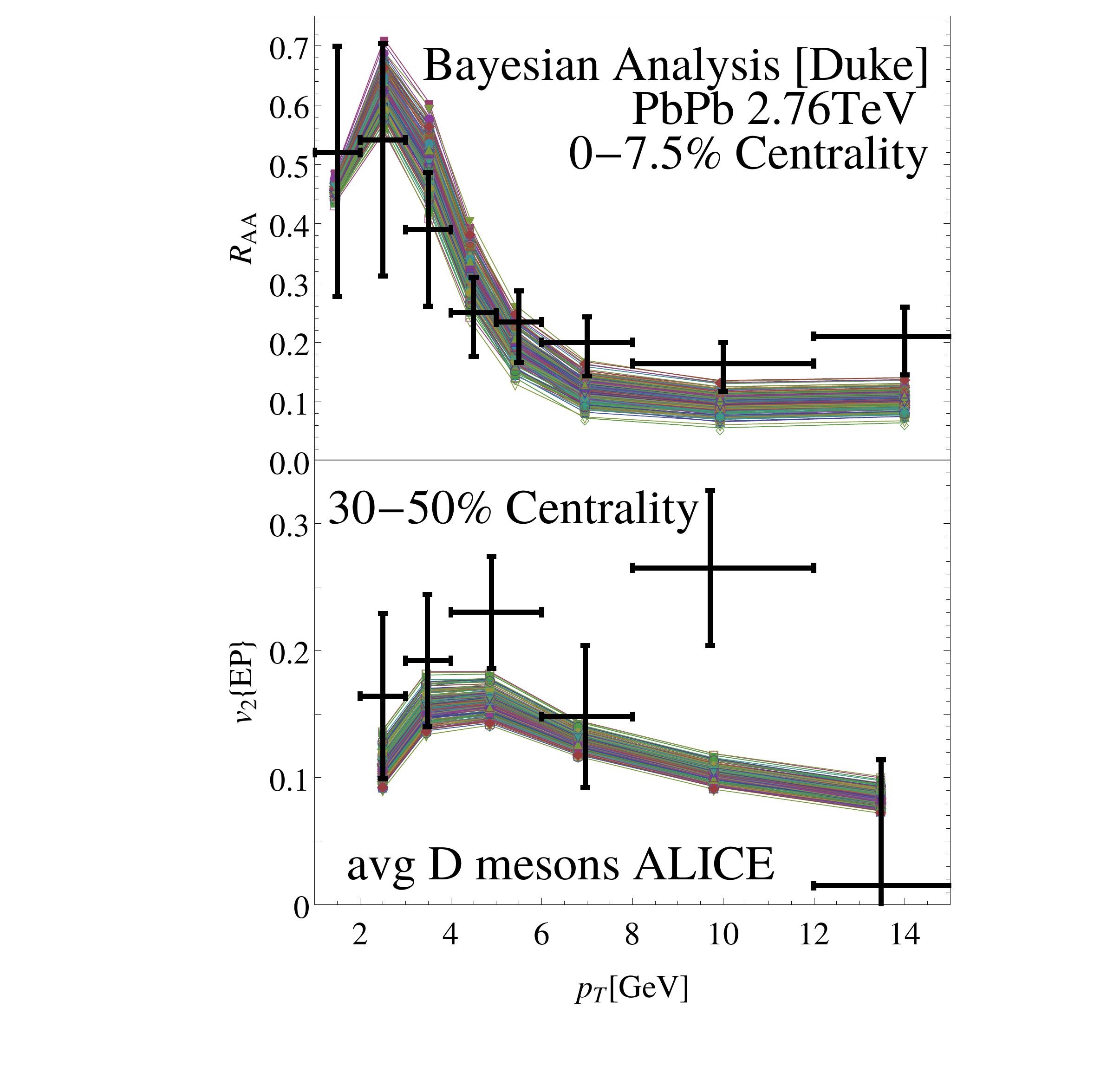}
\caption{(Color online) $R_{AA}$ for $0-7.5\%$ (top) and $v_2\{SP\}$ for $30-50\%$ (bottom) D mesons at LHC PbPb $\sqrt{s_{NN}}=2.76$ TeV using Bayesian Analysis \cite{YingruXU}.}
\label{fig:bay}
\end{figure}

Of course, one could also argue that Fig.\ \ref{fig:SHEE} may be affected by other factors such as initial conditions, viscosity, and the decoupling temperature (where the high $p_T$ particle ceases to interact with the QGP medium).  Thus, the ideal scenario is where an energy loss model coupled to event-by-event hydrodynamics is able to reproduce both soft and hard observables.  There has been a significant advancement in recent years in the soft sector in terms of establishing observables that can more cleanly distinguish between initial conditions and viscosity, see \cite{Zhou:2016eiz} and the references within for a review. A very promising way to explore this is through Bayesian techniques that simultaneously match soft \cite{Bernhard:2016tnd,Pratt:2015zsa} and hard observables and current efforts are already underway in the heavy flavor sector \cite{YingruXU}, as seen in Fig.\ \ref{fig:bay}. 

\subsection{Multiparticle Cumulants}
\label{Multiparticle Cumulants}

\begin{figure}[ht]
\includegraphics[width=0.35\textwidth]{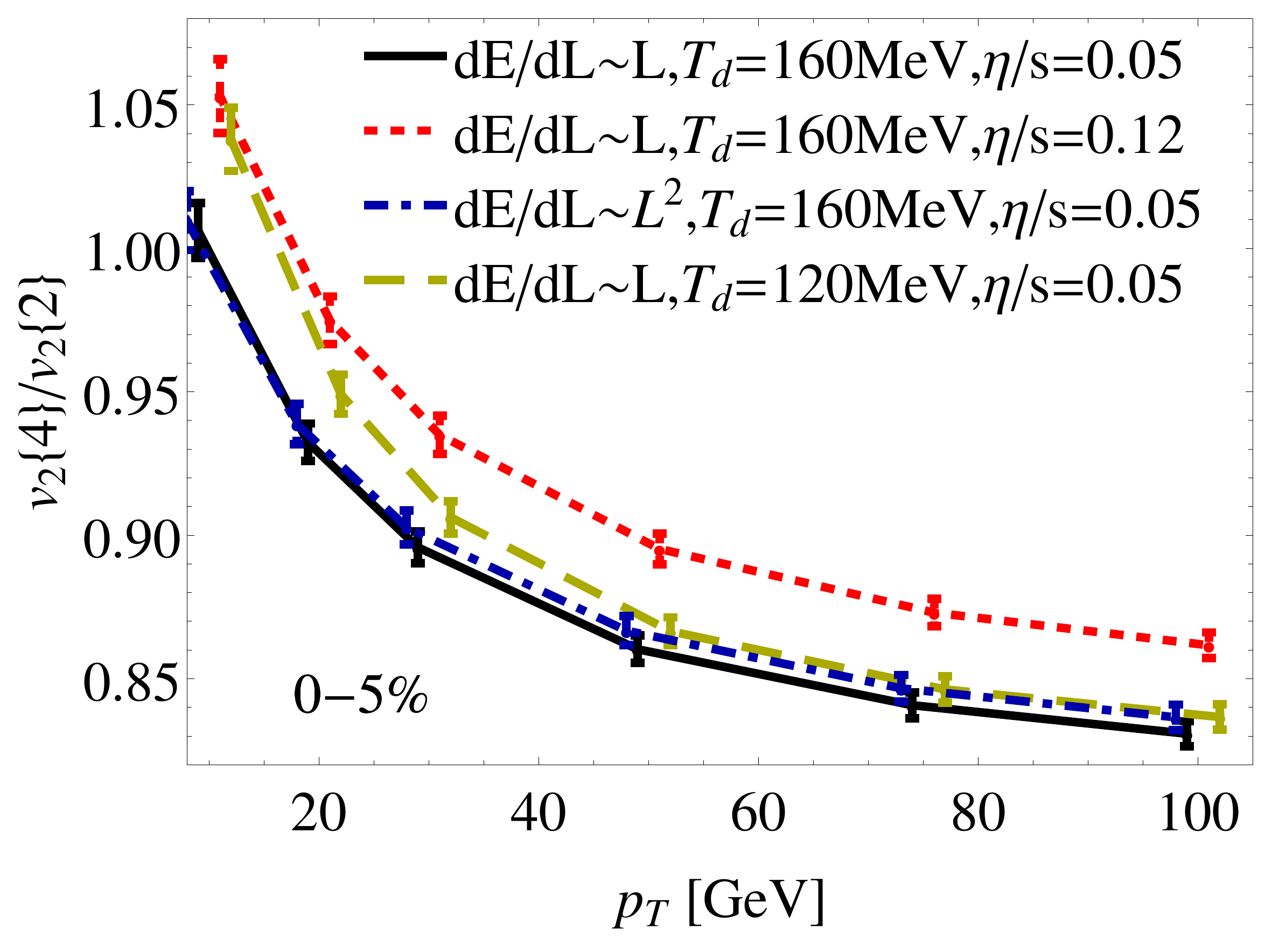} \\
\includegraphics[width=0.38\textwidth]{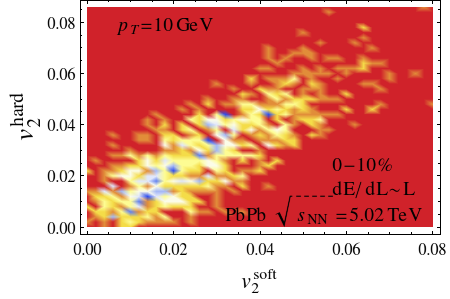}
\caption{(Color online) Predictions for $v_2\{4\}(p_T)/v_2\{2\}(p_T)$ at LHC run 2 (top) and a scatter plot of $v_2^{soft}$ vs. $v_2^{hard}$ on an event-by-event basis (bottom).}
\label{fig:v2cumu}
\end{figure}

Another recent advancement of high $p_T$ flow harmonics was the first calculation \cite{Betz:2016ayq} and measurements \cite{CMS:2016uwf} of multi-particle cumulants. In the soft sector, multiparticle cumulants of flow harmonics integrated over $p_T$ (where 2+ particles are correlated within the same $p_T$ range) are directly related to the moments of the $v_n$ distribution on an event-by-event basis such that
\begin{equation}\label{eqn:kurmom}
\left(\frac{v2\{4\}}{v2\{2\}}\right)^4 = 2 - \frac{\langle v^4 \rangle}{\langle v^2\rangle^2}.
\end{equation}
i.e.  $\left(\frac{v_2\{4\}}{v_2\{2\}}\right)^4$ can give direct information about the kurtosis over the variance of the distribution. In Eq.\ (\ref{eqn:kurmom}), one can see that if there are no event-by-event fluctuations then $\langle v_n^2\rangle^2=\langle v^4 \rangle$, which implies that $v_2\{4\}/v_2\{2\} \rightarrow 1$.  

However, due to limited statistics in the high $p_T$ sector it is not possible to correlate two high $p_T$ particles but rather one must correlate one high $p_T$ particle with one soft particle (or 3 soft particles with 1 high $p_T$ particle for a 4 particle cumulant and so forth). Thus, the direct connection between the kurtosis over variance of the distribution is no longer clear
in the high $p_T$ sector and the relationship $\frac{v_2\{4\}(p_T)}{v_2\{2\}(p_T)}$ actually gives an indication to the degree that linear scaling between the soft and hard sector holds, as demonstrated in \cite{Betz:2016ayq}. The implication of this result is that when $\frac{v_2\{4\}(p_T)}{v_2\{2\}(p_T)}\rightarrow 1$ this does not, in fact, imply that there are no high $p_T$ fluctuations.

In Fig.\ \ref{fig:v2cumu} the ratio $\frac{v_2\{4\}(p_T)}{v_2\{2\}(p_T)}$ is plotted for $p_T>10$ GeV in the centrality class of $0-5\%$ and, around $p_T\sim 10$ GeV, this ratio approaches 1. However, experimentally high $p_T$ flow fluctuations have already been measured up to $p_T=15$ GeV \cite{Aad:2015lwa} so there is direct experimental proof that when $\frac{v_2\{4\}(p_T)}{v_2\{2\}(p_T)}=1$ there are still event-by-event fluctuations.  Additionally, in our calculations the effect of event-by-event fluctuations are always taken into account so we know that $v_2^{hard}$ is still fluctuating.  What is interesting, however, is how much $\frac{v_2\{4\}(p_T)}{v_2\{2\}(p_T)}$ varies from the integrated values that are $\frac{v_2\{4\}}{v_2\{2\}}< 0.8$ for $0-5\%$.  It appears that at around $p_T\sim 10$ GeV there is a maximal divergence from $\frac{v_2\{4\}}{v_2\{2\}}$ and then at higher $p_T$ it relaxes back to $\frac{v_2\{4\}}{v_2\{2\}}$ close to the integrated limit. 

Also shown in Fig.\ \ref{fig:v2cumu} is a scatter plot of $v_2^{soft}$ vs. $v_2^{hard}$ for the $0-10\%$ centrality class   (a wider centrality class is shown to enhance the statistics).  The scatter plot demonstrates not only that there is a strong linear mapping between $v_2^{soft}$ and $v_2^{hard}$ but that even for one specific $v_2^{soft}$ there are fluctuations in the possible $v_2^{hard}$.  This implies that events with small  integrated $v_2^{soft}$ are more likely to produce $v_2^{hard}$ that are small as well (or, conversely, events with a large $v_2^{soft}$ are more likely to produce a large $v_2^{hard}$). Due to non-linear response between $\varepsilon_2\rightarrow v_2$ in the soft sector this relationship is slightly more complicated for peripheral collisions \cite{Noronha-Hostler:2015dbi}.   

Referring back to Eq.\ (\ref{eqn:vncor}), one can see that this relationship also implies that the event plane angles must be strongly correlated in  order to have such a strong linear mapping between soft and hard $v_2$. In fact, one expects that for $v_2$ there is a very strong probability that high $p_T$ particles are emitted in alignment with the soft event plane $\psi_2$ angle.  However, as shown in  \cite{Betz:2016ayq,Jia:2012ez}, that relationship does not hold as strongly for higher order event plane angles. 

\section{Conclusions and Outlook}
\label{Conclusions and Outlook}

In conclusion, it was found that a combination of contributing factors lead to the solution of the $R_{AA}\otimes v_2$ puzzle.  Special attention must be paid to choosing reasonable initial conditions that are able to reproduce the flow harmonics in the soft sector.  Additionally, there is clear experimental evidence that event-by-event fluctuations are influential even up to high $p_T$ and are necessary to take into account in order to do apples-to-apples comparisons between theory and experimental data. In fact, experimental results from ATLAS have found positive $v_3$ results up to $p_T\sim 25$ GeV in a range of centrality classes \cite{ATLAS-CONF-2016-105}. Furthermore, CMS have recently measured multi-particle cumulants up until $p_T\sim 80$ GeV, which appear to have a strong $p_T$ dependence \cite{CMS:2016uwf}.    Furthermore, when one calculates multi-particle cumulants theoretically (especially for 4+ particle calculations) effects such as centrality rebinning and multiplicity weighing play a role and also need to be included.  

While the inclusion of event-by-event fluctuations in jet quenching calculations increases computational costs, it is beyond doubt that they are required to describe current experimental data. In fact, event-by-event fluctuations can actually become a strong asset when one utilizes event shape engineering. As was discussed here, observables constructed using Soft Hard Event Engineering have the potential to more cleanly differentiate between energy loss models. Additionally, the study of multiparticle cumulants opens up entirely new research opportunities (both theoretically and experimentally) involving high $p_T$ flow harmonics that have yet to be explored. In the near future, by describing soft and hard flow harmonics (and spectra) simultaneously across all centralities, a much better understanding of the  jet energy loss mechanism in the QGP will be achieved.

\section*{Acknowledgements}

J.N.H would like to Milos Gyulassy, Barbara Betz, Jorge Noronha and Matthew Luzum for collaborations on this topic.  
J.N.H. was supported by the National Science Foundation under grant
no. PHY-1513864 and acknowledges the use of the Maxwell
Cluster and the advanced support from the Center of Advanced
Computing and Data Systems at the University
of Houston to carry out the research presented here. 

\nocite{*}
\bibliographystyle{elsarticle-num}
\bibliography{library}



\end{document}